\documentclass[conference]{IEEEtran}

\usepackage{tikz}
\usepackage{amssymb}
\usepackage{amsmath}
\usepackage{amsfonts}
\usepackage{mathtools}
\usepackage{mathpartir}
\usepackage{caption}
\usepackage{subcaption}
\usepackage{listings}
\usepackage{multirow}
\usepackage{float}
\usepackage{tabularray}
\usepackage{xurl}

\usepackage[ruled,linesnumbered]{algorithm2e}
\usepackage{threeparttable}
\usepackage{pifont}
\usepackage{bbding}
\usepackage{pdfpages}
\usepackage{xspace}
\usepackage{booktabs}
\usepackage{tikz}
\usepackage{hyperref}
\hypersetup{
  colorlinks   = true,    
  urlcolor     = olive,    
  linkcolor    = red,    
  citecolor    = red      
}

\usepackage{tcolorbox}
\usepackage{tabularx}
\usepackage{makecell}

\usepackage{tikz}
\usetikzlibrary{arrows.meta,positioning}
\usepackage{booktabs}

\usepackage{longtable}

\newcommand{\bheading}[1]{{\vspace{2pt}\noindent{\textbf{#1}}}}

\newcommand{\failurecondition}[1]{%
  \par\addvspace{.5\baselineskip}\begingroup\setlength{\fboxsep}{5pt}\noindent
  \colorbox{red!4}{\parbox{\dimexpr\linewidth-2\fboxsep\relax}{#1}}%
  \par\endgroup}

\newcolumntype{?}{!{\vrule width 1pt}}

\newcounter{note}[section]

\colorlet{Mycolor1}{green!10!orange!90!}

\newcommand{\secref}[1]{\mbox{Sec.~\ref{#1}}\xspace}

\newcommand{\figref}[1]{\mbox{Fig.~\ref{#1}}}

\newcommand{\ignore}[1]{}

\newcommand{\ie}{\textit{i.e.}\xspace}

\newcommand{\tabincell}[2]{\begin{tabular}{@{}#1@{}}#2\end{tabular}}

\newcounter{packednmbr}

\newenvironment{packeditemize}{
\begin{list}{$\bullet$}{
\setlength{\labelwidth}{0pt}
\setlength{\itemsep}{2pt}
\setlength{\leftmargin}{\labelwidth}
\addtolength{\leftmargin}{\labelsep}
\setlength{\parindent}{0pt}
\setlength{\listparindent}{\parindent}
\setlength{\parsep}{1pt}
\setlength{\topsep}{1pt}}}{\end{list}}

\begin{document}

\title{Safe to Resume? Breaking Execution Continuity of Agent Execution via Rollback}

\author{
	\IEEEauthorblockN{
		Guanlong Wu\IEEEauthorrefmark{1}, 
		Dahui Li\IEEEauthorrefmark{1}, 
		Ke Jiang\IEEEauthorrefmark{1},
		Jianyu Niu\IEEEauthorrefmark{2}, 
        Cong Wang\IEEEauthorrefmark{2}, 
        Yinqian Zhang\IEEEauthorrefmark{1},
} 
\IEEEauthorblockA{\IEEEauthorrefmark{1}Southern University of Science and Technology
, \IEEEauthorrefmark{2}City University of Hong Kong}

}

\maketitle

\begin{abstract}
AI agents are moving toward persistent, stateful execution across various applications, accumulating execution state and external effects that are costly to reconstruct after failures.
Checkpoint and rollback (C/R) are becoming essential for recovery, yet their security implications remain largely unexplored. Correct rollback does not imply secure recovery: a faithfully restored checkpoint may resume an execution whose states, assumptions, and external effects never coexisted in any valid history. In this paper, we present the first systematic security study of checkpoint and rollback in existing agent systems. By examining representative agent C/R systems, we characterize the design space of existing C/R mechanisms and develop a general execution model that captures their recovery boundaries and state dependencies. From this model, we identify five fundamental failure modes spanning incomplete or inconsistent internal state, stale external dependencies, nondeterministic replay, and unrecorded external effects. 
We further demonstrate their security impact through three end-to-end attacks on Hermes, Cline, and LangGraph, enabling malware-verification bypass, unauthorized mail forwarding, and double payment.
To systematically study these failures in practice, we develop a multi-agent analysis pipeline that reconstructs execution semantics, identifies violations of the five failure conditions, and validates them through actual rollback.
Across five representative frameworks, our evaluation shows that these failures recur across heterogeneous C/R designs and stem from a common gap between the state restored by a checkpoint and the dependencies required for secure continuation.
\end{abstract}

\section{Introduction}
\label{sec:introduction}
AI agents are moving from short-lived interactions toward persistent,
stateful execution across autonomous research~\cite{schmidgall2025agent},
coding~\cite{yang2024swe}, and enterprise services~\cite{microsoft2025worktrend}.
This shift is already visible in production: Tata Steel reports deploying over
300 specialized agents, while Cognizant operates more than 200 agents serving
350,000 employees~\cite{tatasteel2026agents,cognizant2026multiagent}.
Yet reliability remains a major concern: a 2026 enterprise survey found that
79\% of respondents had reversed an agent action and 42\% reported revenue loss
from agent failures~\cite{koreai2026survey}.
As long-running executions accumulate state and external effects, restarting
from scratch becomes increasingly costly. Checkpoint and rollback (C/R) are
therefore becoming a basic recovery mechanism, with modern systems checkpointing
framework state~\cite{langgraph2026persistence}, workspaces~\cite{cline2026checkpoints},
and execution environments~\cite{e2b2026snapshots}.

C/R periodically records selected execution state and restores a prior
checkpoint after interruption, which has long been studied
for reliable recovery and state continuity maintaince~\cite{chandy1985distributed,elnozahy2002survey,parno2011memoir,
ristenpart2010randomness}.
Recent systems have begun optimizing C/R specifically for agent execution:
CRAB improves recovery through semantics-aware checkpointing~\cite{wu2026crab}, while DeltaBox
provides low-latency sandbox checkpoint and rollback
~\cite{dong2026deltabox}.
Concurrently, recent studies have begun examining the semantics and security of
agent recovery, including valid resume behavior, recovery consistency, and
semantic rollback attacks
~\cite{khan2026resume,yang2026dart}.
These advances establish both the practical importance and security relevance
of agent recovery, but leave a broader question open:
\emph{when rollback carries forward a security-relevant fact, what dependencies
must remain valid for that fact to be safely reused?}

This question is fundamental because recovery carries forward security-relevant facts established earlier in execution: an artifact has been validated, an operation authorized, a resource binding remains valid, or an external effect has not yet occurred. Yet the validity of these facts may depend on state distributed across framework components, files and runtimes, nondeterministic decisions, and external services, which do not necessarily share the same recovery scope. Thus, a checkpoint's recovery boundary may not cover the full set of security dependencies on which the restored state relies. \emph{Correct rollback does not imply secure recovery}: every recorded state may be restored faithfully while the relationships that made those states security-valid are broken.

Consider a coding agent tasked with removing malware from a repository before release. The agent removes the malicious payload, scans the cleaned repository, and records the passing result. If recovery restores the pre-cleanup workspace while retaining the post-scan agent state, the agent may release the malicious repository as verified. Neither the checkpoint nor the scan result is forged; rollback has rebound a genuine security judgment to an artifact it never validated. We call the underlying requirement \emph{execution continuity}: security-relevant states, decisions, assumptions, and effects carried across recovery must remain jointly consistent with the valid execution history that justifies them.

In this paper, we conduct a systematic cross-system security study of
checkpoint and rollback in agent systems. We first examine representative
agent C/R systems and characterize their designs by three recurring recovery
boundaries: framework state, workspace state, and OS/VM state. This study
motivates a general execution model that captures independently evolving
internal states, nondeterministic execution, checkpoints with partial recovery
boundaries, and an outside world that does not roll back with the agent.

Using this model, we identify five recurring ways in which rollback can break
execution continuity: \emph{incomplete internal state coverage},
\emph{inconsistent checkpoint state}, \emph{external state mismatch},
\emph{unbound nondeterministic replay}, and \emph{unrecorded external effects}.
These failures capture how a legitimately restored checkpoint can invalidate
the state, context, or effects that justify security-relevant decisions carried
across recovery. Importantly, they do not require checkpoint corruption or a
compromised recovery mechanism.

We demonstrate the security impact of these failures through three end-to-end
attacks on Hermes, Cline, and LangGraph, causing malware-verification bypass,
unauthorized mail forwarding, and duplicate payment, respectively. We further
develop a trace-based framework that reconstructs execution dependencies,
detects candidate recovery failures, and validates them through native rollback
and re-execution. Across 347 benchmark traces from TerminalBench and AgentBench,
yielding 1,735 framework-task executions over five representative C/R systems,
our evaluation shows that recovery failures are common but exhibit distinct
profiles across different recovery boundaries, and reveals recurring
factors involving external dependencies, nondeterministic replay, and persistent
effects.

\bheading{Contributions.} Main contributions are listed as follows:
\begin{packeditemize}

    \item \textbf{Agent recovery abstraction.}
    We characterize representative agent C/R systems and develop a 
    execution model that captures heterogeneous recovery boundaries,
    nondeterministic execution, and an independently evolving outside world.

    \item \textbf{Security characterization and attacks.}
    We introduce execution continuity and identify five recurring ways in which
    legitimate rollback can invalidate security-relevant state or transitions.
    We demonstrate their practical impact through three end-to-end attacks on
    real agent frameworks.

    \item \textbf{Systematic detection and validation.}
    We develop and evaluate a trace-based methodology for detecting recovery
    failures and validating them through actual rollback, providing an empirical
    characterization of how these failures manifest across representative agent
    C/R designs.
\end{packeditemize}

\begin{table*}[t]
\centering
\setlength{\tabcolsep}{3.5pt}
\caption{Checkpoint and rollback mechanisms in representative agent systems.}
\label{tab:checkpoint_systems}
\begin{tabularx}{\textwidth}{
    @{}p{0.080\textwidth}
    p{0.105\textwidth}
    p{0.255\textwidth}
    p{0.245\textwidth}
    X@{}}
\toprule
\textbf{System} &
\textbf{State Scope} &
\textbf{Captured State} &
\textbf{Checkpoint Creation} &
\textbf{Restoration} \\
\midrule

LangGraph
& Framework
& Graph state, including messages, intermediate results, pending execution state
& Automatic at each superstep boundary once persistence is enabled
& Resume the latest checkpoint or select an earlier checkpoint for replay/branching
\\

CrewAI
& Framework
& Crew/Flow/Agent execution state, including task progress, outputs, memory, and intermediate state
& Event-driven once enabled; task completion is the default trigger
& Restore a selected checkpoint and resume without re-executing completed tasks
\\

Hermes
& Workspace
& Project/workspace files; rollback also adjusts the most recent conversation turn
& Automatic before file mutations or recognized destructive commands once checkpointing is enabled
& Restore a selected workspace checkpoint (or individual file) via \texttt{/rollback}
\\

Cline
& Workspace
& Project files associated with the corresponding task-history position
& Automatic after file modifications and terminal commands; enabled by default
& Restore files, task history, or both from a selected checkpoint
\\

E2B
& OS/VM
& Sandbox filesystem and memory state, including running processes
& Explicit, at developer-selected points via \texttt{create\_snapshot()}
& Create a new sandbox from a selected snapshot
\\

\bottomrule
\end{tabularx}
\end{table*}

\section{Characterizing Existing Agent Checkpointing}
\label{sec:background}

We examine 12 representative agent C/R systems and classify their checkpointing mechanisms by \emph{recovery boundary}, \ie, the primary state that is restored upon rollback. As summarized in Table~\ref{tab:checkpoint_systems} for five representative systems, existing designs fall into three recurring categories: \emph{framework-state}, \emph{workspace-state}, and \emph{OS/VM-state} checkpoints. For each category, we characterize \emph{what state is captured}, \emph{when checkpoints are created}, and \emph{how state is restored}.

\subsection{Framework-State Checkpoints}
\label{sec:bg_agent_state}

Framework-state checkpoints preserve framework-managed execution state, such as conversation history, workflow progress, task outputs, agent memory, and pending steps, while leaving workspace and runtime state outside the recovery boundary. Representative systems include LangGraph~\cite{langgraph2026persistence}, CrewAI~\cite{crewai_overview}, LlamaIndex Workflows~\cite{llamaindex_workflows}, Google ADK~\cite{google_adk}, and Microsoft Agent Framework~\cite{microsoft_agent_framework}. In this paper, we focus on LangGraph and CrewAI.

\bheading{LangGraph.}
LangGraph represents an agent workflow as a graph whose nodes perform operations such as LLM calls or tool invocations.
Execution proceeds in \emph{supersteps}: each superstep executes the current scheduled nodes before advancing to the next graph state, thereby providing a checkpoint boundary.

\begin{packeditemize}

\item \textit{Checkpointing.}
When persistence is enabled, LangGraph associates each execution history with a \texttt{thread\_id} and uses a configured checkpointer to save a \texttt{StateSnapshot} at every superstep boundary.
A snapshot records the accumulated graph state, such as conversation messages and intermediate results, together with metadata describing the nodes and pending tasks scheduled for subsequent execution.
Snapshots can be persisted in backends such as SQLite or PostgreSQL for durable recovery, and LangGraph does not impose a default retention limit on the checkpoint history~\cite{langgraph2026persistence}.

\item \textit{Restoration.}
An application may resume from the latest checkpoint using the same \texttt{thread\_id}, or select an earlier checkpoint from the state history for replay or branching.
Execution resumes from the corresponding superstep boundary: state represented before that checkpoint is retained, whereas subsequent nodes, including their LLM calls and tool invocations, are re-executed~\cite{langgraph2026persistence,langgraph_timetravel}.

\end{packeditemize}

\bheading{CrewAI.}
CrewAI organizes applications around two execution abstractions: a \emph{Crew} coordinates role-specific agents and tasks, whereas a \emph{Flow} maintains shared state and orchestrates event-driven execution involving code, LLM calls, and Crews~\cite{crewai_overview}.

\begin{packeditemize}

\item \textit{Checkpointing.}
CrewAI checkpoints framework-managed execution state.
For Crews, this includes configuration and inputs, task progress and outputs, and participating agents' memory and runtime attributes; for Flows, it includes shared state, completed methods, intermediate outputs, and execution position.
Checkpoint creation is event-driven: task completion (\texttt{task\_completed}) is the default trigger, while finer-grained events such as completed LLM calls or Flow methods can also be configured; developers may additionally create checkpoints explicitly.
Checkpoints can be persisted using JSON or SQLite backends, with \texttt{max\_checkpoints} optionally bounding the retained checkpoint history~\cite{crewai_checkpointing}.

\item \textit{Restoration.}
A stored checkpoint can be selected to restore the corresponding framework state and resume execution from that point, without re-executing tasks recorded as completed in the checkpoint~\cite{crewai_checkpointing}.

\end{packeditemize}

\subsection{Workspace-State Checkpoints}
\label{sec:bg_agent_fs}

Workspace-state checkpoints extend recovery to artifacts manipulated by the agent, most commonly project files.
They are particularly common in coding agents and are typically implemented using file snapshots or a \emph{shadow Git repository}---a separate repository used to version the agent's workspace without modifying the project's own Git history.
Their recovery boundary, however, generally excludes runtime state such as process memory, installed system state, active services, and files outside the selected workspace.
Representative systems include Hermes~\cite{hermes_checkpointing}, Cline~\cite{cline_overview}, Gemini CLI~\cite{gemini_checkpointing}, and Claude Code~\cite{claude_checkpointing}; we focus on Hermes and Cline.

\bheading{Hermes.}
Hermes is a general-purpose tool-using agent that follows an iterative tool-calling loop, allowing it to manipulate files and execute terminal commands while maintaining the surrounding conversation state~\cite{hermes_agent}.

\begin{packeditemize}

\item \textit{Checkpointing.}
When checkpointing is enabled, Hermes creates a workspace checkpoint immediately before file-writing operations such as \texttt{write\_file} and \texttt{patch}, as well as recognized destructive terminal commands.
It resolves the relevant project root, creates at most one checkpoint per directory within each conversation turn, and stores workspace snapshots as project-specific references in a shared shadow Git store.
By default, Hermes retains at most 20 checkpoints per project and 500\,MB across the shared store, while excluding individual files larger than 10\,MB~\cite{hermes_checkpointing}.

\item \textit{Restoration.}
Users invoke \texttt{/rollback} to select a previous checkpoint.
Hermes restores the files represented by the selected Git tree and rolls back the most recent conversation turn.
Its recovery restores workspace state together with only a limited portion of the surrounding conversation state~\cite{hermes_checkpointing}.

\end{packeditemize}

\bheading{Cline.}
Cline is an IDE-integrated coding agent in which each task maintains a user goal, conversation, and task history, and a sequence of tool interactions over a project workspace.
During execution, Cline can inspect and modify files, execute terminal commands, and invoke browser or Model Context Protocol (MCP) tools, subject to explicit user approval or configured auto-approval policies~\cite{cline_overview,cline_tasks}.

\begin{packeditemize}

\item \textit{Checkpointing.}
Cline enables checkpoints by default and maintains a shadow Git repository independent of the project's Git history.
It snapshots the complete file state of the current workspace, including files not tracked by the user's repository, and associates each snapshot with the corresponding position in the task history.
Checkpoints are created automatically after relevant tool use, including file modifications and terminal commands, and persist across editor sessions; Cline has no default retention bound~\cite{cline_checkpoints}.

\item \textit{Restoration.}
Cline allows the user to restore the workspace while retaining the current task history, restore the task history while retaining the current workspace, or restore both to the selected checkpoint.
Thus, workspace state and framework-managed state can be handled independently~\cite{cline_checkpoints}.

\end{packeditemize}

\subsection{OS/VM-State Checkpoints}
\label{sec:bg_os_vm}

OS/VM-state checkpoints extend the recovery boundary to the execution environment, capturing filesystems, process memory, and running services at higher checkpointing and storage costs.
Representative systems include E2B~\cite{e2b_snapshots}, CRAB~\cite{wu2026crab}, and DeltaBox~\cite{dong2026deltabox}.
While CRAB and DeltaBox reduce these costs through semantics-aware or incremental checkpointing, we use E2B as a representative system.

\bheading{E2B.}
E2B provides isolated cloud \emph{sandboxes}---self-contained execution environments in which agents can manipulate files, execute programs, and run services.
Each sandbox maintains its own filesystem, processes, and memory state and is accessed through the E2B SDK.

\begin{packeditemize}

\item \textit{Checkpointing.}
An application explicitly invokes \texttt{create\_snapshot()} to create a point-in-time snapshot of a running sandbox.
E2B briefly pauses the sandbox while capturing its filesystem and memory state, including running processes and loaded data, after which the original sandbox continues execution.
Snapshots persist until deleted, with no default retention period or history bound~\cite{e2b_snapshots}.

\item \textit{Restoration.}
A snapshot identifier can be supplied to \texttt{Sandbox.create()} to instantiate a new sandbox from the captured filesystem and memory state.
Snapshots are reusable, allowing the same captured state to initialize multiple sandboxes~\cite{e2b_persistence,e2b_snapshots}.

\end{packeditemize}

\failurecondition{\textcolor{red!75!black}{\textbf{Takeaway.}}~
Agent C/R mechanisms span framework, workspace, and runtime state, but broader coverage does not guarantee safe recovery.
External state and persistent effects remain outside the recovery boundary, while some designs restore state components independently.
Thus, recovery may break the dependencies required for security-valid resumption, motivating the general model in \secref{sec:model}.}

\section{Problem Definition}
\label{sec:problem}

We develop a general model for
agent execution and rollback based on the recovery boundaries characterized in
\secref{sec:background}. Our model draws on the classic distributed-state
abstraction of Chandy and Lamport~\cite{chandy1985distributed}, while
capturing three characteristics central to agent recovery: heterogeneous
internal state, nondeterministic execution, and interaction with an external
world that does not roll back with the agent.

\subsection{System Model}
\label{sec:model}

\bheading{Components.}
We model an agent system using the following components.

\begin{packeditemize}

\item \textbf{Process.}
The agent system consists of a set of logical processes
\(\mathcal{P}=\{P_1,P_2,\ldots,P_n\}\).
Each \(P_i\) represents an independently evolving stateful component rather
than necessarily an operating-system process~\cite{chandy1985distributed}.
For example, the main agent, a subagent, framework-managed conversation
state, a tool runtime, or the workspace filesystem may each be modeled as a
process.

\item \textbf{Process state.}
Each process \(P_i\) evolves through a sequence of local states
\(s_i^0,s_i^1,\ldots\), where \(s_i^j\in\mathcal{S}_i\) denotes the state
after its \(j\)-th local transition and \(\mathcal{S}_i\) is its state space.
Depending on the process, this state may include conversation history,
workflow progress, intermediate results, permissions, in-memory variables,
or filesystem contents.
A global internal state combines one local state from each process:
\[
S=(s_1^{k_1},s_2^{k_2},\ldots,s_n^{k_n}).
\]
The indices \(k_i\) need not be equal because different processes evolve
independently.

\item \textbf{Event.}
An event \(e=(a,\xi)\) represents an operation \(a\) together with the
nondeterministic factors \(\xi\) that determine its realized outcome,
inducing a transition
\[
s_i^j \xrightarrow{e} s_i^{j+1}.
\]
Events include internal computation, agent decisions, and tool invocations,
while \(\xi\) may capture LLM sampling, randomness, or responses from
external services.

\item \textbf{Message.}
A message \(m\) carries information to a process from another process or
from the outside world.
Examples include user inputs, inter-agent messages, and tool results.
A delivered message may trigger an event and thereby change the receiving
process state.
We write \(e_a \prec e_b\) when \(e_a\) causally precedes \(e_b\), either
through local execution order or through such an interaction.

\item \textbf{Checkpoint.}
A checkpoint records selected process states as a recovery point.
Let \(K\subseteq\mathcal{P}\) denote the processes covered by a checkpoint.
We write
\[
C_K=(s_i^{k_i})_{P_i\in K}.
\]
The set \(K\) therefore defines the checkpoint's \emph{recovery boundary}.
A full-state checkpoint is the special case \(K=\mathcal{P}\).
During execution, the system may retain multiple checkpoints,
\(\mathcal{C}=\{C^1,\ldots,C^q\}\).
We do not model in-flight messages as a checkpoint component.
When a framework persists pending work or communication state, such
information is included in the corresponding process state.
This abstraction captures the coarse execution boundaries used by the
agent systems characterized in \secref{sec:background}.

\item \textbf{Outside world.}
The outside world \(W\) contains state and entities beyond the agent system
that are not directly restored by agent rollback.
Examples include remote services, payment systems, mailboxes, databases,
user or organizational state, and persistent effects of previous agent
actions.
Processes may read from or modify \(W\), but restoring an agent checkpoint
does not in general restore the corresponding external state.
Thus, even a checkpoint with \(K=\mathcal{P}\) need not restore the complete
state on which the execution depends.

\end{packeditemize}

\begin{figure}[t]
\centering
\includegraphics[width=\columnwidth]{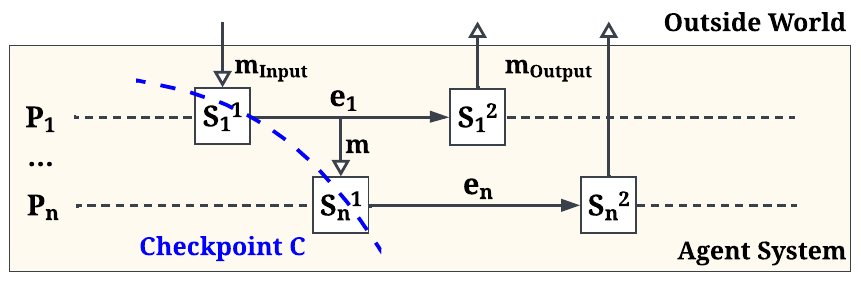}
\caption{System model for agent checkpoint and rollback.}
\label{fig:system_model}
\end{figure}

\bheading{Execution and recovery.}
Execution begins when an internal process receives an input, for example a
user instruction delivered to the main agent.
The resulting events update local states and may exchange messages or
interact with \(W\), producing further state transitions and external
effects.
During execution, the system records selected process states as checkpoints
\(C\in\mathcal{C}\).

Consider rollback at time \(t_r\) to a checkpoint \(C_K\) created earlier.
For every \(P_i\in K\), recovery restores the state recorded in \(C_K\).
Processes outside \(K\) are not controlled by this checkpoint and may retain
their current state, restart from an earlier state, or be reconstructed from
another source; we denote the recovered state of such a process \(P_j\) by
\(\rho_j\).
Meanwhile, the outside world remains at its current state \(W^{t_r}\).
Rollback therefore reconstructs the configuration
\[
\widehat{X}
=
\left(
(s_i^{k_i})_{P_i\in K}
\oplus
(\rho_j)_{P_j\notin K},
W^{t_r}
\right),
\]
from which execution resumes.

This formulation makes explicit a key property of C/R: a recovered
configuration may combine checkpointed state, uncovered internal state, and
external state originating from different points in the execution.

\subsection{Execution Continuity}
\label{sec:execution-continuity}

Correctly restoring the contents of a checkpoint does not by itself imply
that the resulting execution is security-valid.
We capture this distinction through \emph{execution continuity}: security-
relevant states, decisions, assumptions, and effects carried across recovery
must remain jointly consistent with a valid continuous execution history.

We use a \emph{valid continuous execution} to denote an execution that
follows the agent workflow and its security constraints without rollback,
under the corresponding inputs, authorizations, and external interactions.
Recovery can violate execution continuity in two fundamental ways~\cite{parno2011memoir}:

\begin{packeditemize}

\item \textbf{Invalid recovered state.}
Rollback reconstructs a combination of states or assumptions that could not
jointly occur in a valid continuous execution.
For example, an agent may retain a successful validation result while
restoring the validated artifact to an earlier, unverified version.

\item \textbf{Invalid recovery transition.}
The recovered state may itself be individually valid, but rollback enables
a subsequent transition that no valid continuous execution would permit.
For example, restoring a pre-payment state may allow a one-time approval to
authorize a payment whose external effect has already occurred.

\end{packeditemize}

The security failures in \secref{sec:security-failures} characterize
different ways in which C/R breaks these two forms of execution continuity.

\subsection{Threat Model}
\label{subsec:threatmodel}

\bheading{Adversary goal.}
The adversary aims to exploit C/R to violate execution continuity, causing
the agent either to resume from an invalid recovered state or to perform an
invalid recovery transition.
Importantly, the adversary need not corrupt a checkpoint or compromise the
recovery implementation; the attack instead exploits how legitimate rollback
behavior interacts with agent state and the outside world.

\bheading{Adversary capabilities.}
We assume the adversary knows the task or workflow and the system's publicly
documented C/R behavior, consistent with malicious skills, MCP servers,
tools, and external services considered in prior agent-security
work~\cite{agentdojo2025,skillinject2026,mcptox2026,greshake2023not}.
Depending on the concrete setting, the adversary may influence agent inputs,
workflow instructions, or tool and service responses.
Such influence can steer execution toward operations that trigger checkpoint
creation or place the system in an advantageous state around a checkpoint.

The adversary may also induce a recoverable failure or interruption, for
example by causing a malicious tool, MCP server, or external service to
return an error or malformed response.
Recovery may then be initiated through the framework's normal mechanism by
the user, framework, or deployment environment.
When a rollback interface is directly exposed to an adversarial component,
the adversary may invoke it within the permissions provided by that
interface.

\bheading{Trusted components.}
The adversary cannot directly modify checkpoint contents or storage, alter
the semantics of the C/R implementation, or forge security decisions
provided by otherwise trusted parties.
We also do not assume deterministic replay: LLM outputs, tool responses, and
other nondeterministic factors may legitimately differ after rollback.
Each attack in \secref{sec:attack-cases} instantiates this general threat
model with a concrete adversary and states its additional capabilities.
\section{Security Failures in Agent C/R}
\label{sec:security-failures}

Under the model in \secref{sec:problem}, we identify five recurring ways
in which checkpoint and rollback can violate \emph{execution continuity}.
The first two concern the internal state reconstructed by recovery:
a checkpoint may omit recovery-relevant state (\textbf{SF1}) or combine
captured states that are mutually inconsistent (\textbf{SF2}).
The remaining failures arise from relations that cross the checkpoint
boundary: restored state may become incompatible with the current outside
world (\textbf{SF3}), nondeterministic replay may follow a security-relevant
execution path different from the original one (\textbf{SF4}), or rollback
may forget effects that have already persisted externally (\textbf{SF5}).
These conditions can yield either an invalid recovered state or an invalid
recovery transition, as defined in \secref{sec:execution-continuity}.

\subsection{SF1: Incomplete Internal State Coverage}
\label{sec:pattern-incomplete}

\begin{figure}[t]
    \centering
    \includegraphics[width=\columnwidth]{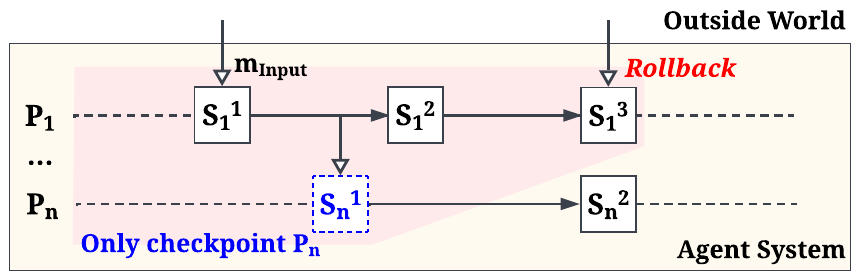}
    \caption{SF1: Incomplete internal state coverage.}
    \label{fig:incomplete_state_coverage}
\end{figure}

\bheading{Definition (\figref{fig:incomplete_state_coverage}).}
Incomplete internal state coverage occurs when a checkpoint omits internal
state required to safely resume the execution.
For a task \(T\), let \(\mathcal{P}_T\subseteq\mathcal{P}\) denote the
internal processes whose states are relevant to the resumed execution, and
let
\[
S_T=(s_i^{k_i})_{P_i\in\mathcal{P}_T}
\]
denote their states when checkpoint \(C_K\) is created.
The checkpoint has incomplete coverage if
\(\mathcal{P}_T\nsubseteq K\).
Upon rollback, states in \(K\) are restored from \(C_K\), whereas an omitted
process \(P_j\notin K\) recovers to some state \(\rho_j\), as modeled in
\secref{sec:model}.
If \(\rho_j\neq s_j^{k_j}\), recovery may combine internal states that never
coexisted at the checkpointed execution point.

\bheading{Consequences.}
The omitted state can diverge from the checkpoint in two directions:

\begin{packeditemize}

\item \emph{Newer-state carryover.}
An uncovered process may remain live across rollback and retain a state
reached after the checkpoint.
Specifically, if \(P_j\) evolves from \(s_j^{k_j}\) at checkpoint creation
to \(s_j^{k'_j}\), where \(k'_j>k_j\), rollback leaves
\(\rho_j=s_j^{k'_j}\).
The recovered configuration therefore combines checkpointed state with a
later state of \(P_j\).

\item \emph{Older-state fallback.}
An uncovered process may instead restart after a failure without access to
its checkpoint-time state.
It may therefore recover from \(s_j^{k'_j}\), where \(k'_j<k_j\), or from
its initial state.
The recovered configuration then combines checkpointed state with an
earlier state of \(P_j\).

\end{packeditemize}

For example, an agent validates a deployment manifest and records
\texttt{validated=true} in framework state, while the manifest resides in an
ephemeral workspace outside the checkpoint.
After a crash, the workspace is reconstructed from its original template
while the framework restores the post-validation state.
The resumed agent may therefore deploy a manifest that was never validated.

\bheading{In existing designs.}
Incomplete coverage can arise both from the intended recovery boundary and
from implementation-specific exclusions.
Framework-state checkpoints omit workspace and runtime state;
workspace-state checkpoints additionally preserve workspace artifacts but
generally omit process and environment state.
Even OS/VM-state checkpoints need not cover framework-managed state stored
outside the snapshotted environment or state maintained by external
services.
Concrete implementations can narrow coverage further: Git-based mechanisms
typically snapshot only a selected project scope, while Hermes, for example,
excludes individual files larger than 10\,MB and applies additional
scope and storage limits~\cite{hermes_checkpointing}.

\subsection{SF2: Inconsistent Checkpoint State}
\label{sec:pattern-inconsistent}

\begin{figure}[t]
    \centering
    \includegraphics[width=\columnwidth]{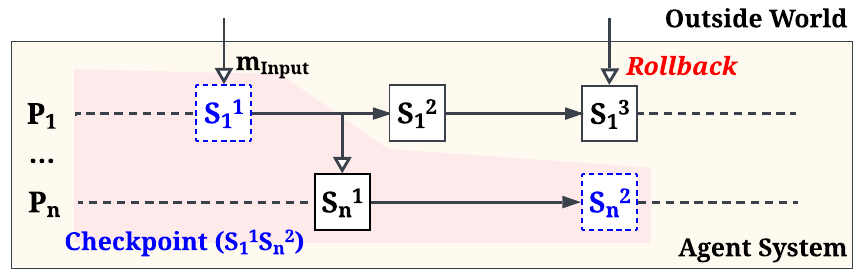}
    \caption{SF2: Inconsistent checkpoint state.}
    \label{fig:causal_consistency}
\end{figure}

\bheading{Definition (\figref{fig:causal_consistency}).}
SF1 concerns state that is \emph{missing}; SF2 concerns captured states that
are individually present but \emph{mutually inconsistent}.
Let
\[
C_K=(s_i^{k_i})_{P_i\in K}
\]
be a checkpoint, and let \(H_i(s_i^{k_i})\) denote the local event history
reflected by state \(s_i^{k_i}\).
Define
\[
H(C_K)=\bigcup_{P_i\in K}H_i(s_i^{k_i}).
\]
Using the causal relation \(\prec\) from \secref{sec:model}, \(C_K\) is
causally consistent if, for every \(e_b\in H(C_K)\), each causal predecessor
\(e_a\prec e_b\) belonging to a captured process also appears in
\(H(C_K)\).
Otherwise, the checkpoint contains states that cannot correspond to a single
causally valid execution cut.

For example, suppose captured state of \(P_n\) reflects consumption of an
event produced by \(P_1\), while the captured state of \(P_1\) precedes the
production of that event.
Both local states may be valid individually, yet they cannot belong to the
same execution history.

\bheading{Consequences.}
Rollback to an inconsistent checkpoint reconstructs an invalid internal
state.
A result, approval, or validation may survive without the state that
produced or justified it; control state may indicate that an operation has
completed while associated data predates the operation; or different
components may disagree on execution progress, causing later actions to be
skipped, repeated, or reordered.

For example, an agent patches a vulnerable file, tests the patched version,
and records \texttt{tests\_passed=true} in its conversation state.
If recovery restores the conversation state from after testing but the
workspace from before the patch, both restored states are individually
valid but do not form a consistent causal cut.
The agent may then submit the vulnerable file without repeating the patch or
tests.

\bheading{In existing designs.}
Framework-defined checkpoints can preserve consistency within the
framework-managed state when that state is captured at a common execution
boundary.
Workspace-state mechanisms introduce an additional state domain whose
checkpoint and restoration may not be coupled to framework state.
Cline provides a concrete example: it allows the workspace, task history,
or both to be restored from a checkpoint, so the two components can
intentionally be rolled back independently~\cite{cline_checkpoints}.
OS/VM-state mechanisms can atomically capture a broader execution
environment, preserving consistency within the recovery boundary.

\subsection{SF3: External State Mismatch}
\label{sec:pattern-external-mismatch}

\begin{figure}[t]
    \centering
    \includegraphics[width=\columnwidth]{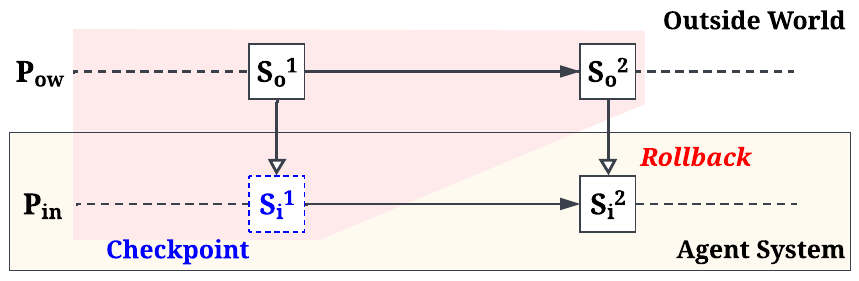}
    \caption{SF3: External state mismatch.}
    \label{fig:external_state_mismatch}
\end{figure}

\bheading{Definition (\figref{fig:external_state_mismatch}).}
External state mismatch occurs when checkpointed internal state is restored
against an outside world under which the assumptions encoded by that state
are no longer valid.
Let \(s_i^{k_i}\) be an internal state captured in checkpoint \(C_K\), and
let \(W^{t_c}\) denote the outside-world state on which it depends when the
checkpoint is created at time \(t_c\).
We define
\[
\Gamma(s_i,W)\in\{0,1\}
\]
as a compatibility predicate indicating whether the security-relevant
judgments, or assumptions represented by \(s_i\) remain
valid under outside-world state \(W\).
At checkpoint creation,
\[
\Gamma(s_i^{k_i},W^{t_c})=1.
\]
Because rollback does not restore the outside world, recovery at
\(t_r>t_c\) combines \(s_i^{k_i}\) with \(W^{t_r}\).
An external state mismatch occurs when
\[
\Gamma(s_i^{k_i},W^{t_r})=0.
\]

\bheading{Consequences.}
A decision or transition that was valid under \(W^{t_c}\) may no longer be
permitted under \(W^{t_r}\).
The agent may reuse a validation, approval, or trust decision for a changed
external object; continue after policy or ownership status has
changed; or act on an external binding that no longer holds.

For example, suppose an agent checkpoints a recipient list used to send
sensitive documents to current employees.
If an employee leaves after the checkpoint is created, rollback may restore
the obsolete recipient list while the organization's membership state
remains current.
The resumed agent can then disclose subsequent documents to a former
employee.

\bheading{In existing designs.}
Among the systems reviewed in \secref{sec:background}, we find no general
mechanism that couples checkpoint restoration with revalidation of all
external dependencies.
This limitation is structural: the relevant outside-world state may be
owned by remote services and remain beyond the recovery boundary even when
the agent's local execution environment is fully snapshotted.
We discuss its relation to traditional stale-state recovery in
\secref{sec:discussion}.

\subsection{SF4: Unbound Nondeterministic Replay}
\label{sec:pattern-nondeterminism}

\begin{figure}[t]
    \centering
    \includegraphics[width=\columnwidth]{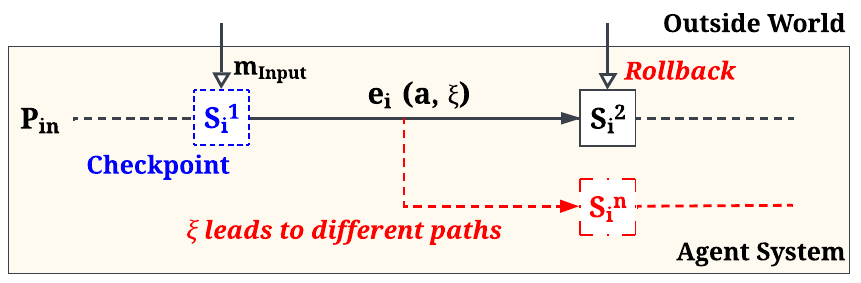}
    \caption{SF4: Unbound nondeterministic replay.}
    \label{fig:unbound_nondeterministic_reexecution}
\end{figure}

\bheading{Definition (\figref{fig:unbound_nondeterministic_reexecution}).}
Rollback commonly re-executes operations that occurred after the selected
checkpoint.
Such replay becomes security-relevant when, under the same restored state and relevant external conditions, a nondeterministic operation produces a different outcome that is not bound to the state or security decisions carried across recovery.

Suppose operation \(a\), executed after checkpoint state \(s_i^j\), realizes
event
\[
e_1=(a,\xi_1)
\]
and transitions to \(s_i^{j+1}\).
After rollback to \(s_i^j\), the same operation may execute as
\[
e_2=(a,\xi_2), \qquad \xi_2\neq\xi_1,
\]
producing a different security-relevant outcome
\(\hat{s}_i^{j+1}\neq s_i^{j+1}\).
An \emph{unbound nondeterministic replay} occurs when recovery permits this
divergent outcome without preserving the security
relationships established for the original execution.

\bheading{Consequences.}
Two forms are particularly security-relevant.

\begin{packeditemize}

\item \emph{Cross-path reuse.}
Security-relevant state established for one continuation---such as an
approval, validation result, permission, or object binding---remains usable
after replay follows a different continuation for which that state is no
longer valid.

\item \emph{Retry amplification.}
Repeated rollback turns a security-sensitive nondeterministic decision into
repeated trials until a favorable outcome appears.

\end{packeditemize}

For example, consider a release agent that checkpoints immediately before a
randomized regression test that samples a subset of the test suite.
The original execution samples a test that detects a defect but fails before
the result is durably recorded.
After rollback, replay draws a different sample that omits the detecting
test, allowing the build to pass and be released.
Neither test execution is individually invalid; the violation arises because
rollback substitutes one security-relevant outcome for another.

\bheading{In existing designs.}
The frameworks reviewed in \secref{sec:background} generally permit
post-checkpoint LLM calls and tool operations to be executed again, but we
find no general mechanism that binds security-relevant nondeterministic
outcomes across replay or constrains repeated rollback.
The issue is particularly salient for agents because model sampling and
external tool responses make nondeterminism a routine part of execution,
rather than an exceptional source of divergence.

\subsection{SF5: Unrecorded External Effects}
\label{sec:pattern-external-effect}

\begin{figure}[t]
    \centering
    \includegraphics[width=\columnwidth]{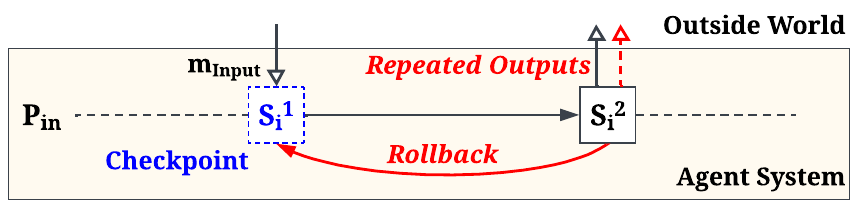}
    \caption{SF5: Unrecorded external effects.}
    \label{fig:uncoordinated_external_effects}
\end{figure}

\bheading{Definition (\figref{fig:uncoordinated_external_effects}).}
An unrecorded external effect occurs when an action executed after a checkpoint produces an effect outside the recovery boundary, but rollback to that checkpoint removes the internal record that the action has occurred.
Specifically, let checkpoint $C$ capture internal state $s_i^j$.
After $C$, an action $a$ transitions the internal state and produces an external effect:
\[
s_i^j \xrightarrow{a} s_i^{j+1},
\qquad
W \xrightarrow{a} W'.
\]
If the system later rolls back to $C$, it restores $s_i^j$, while the external effect persists in $W'$.
The recovered execution therefore reaches $a$ again from a state in which the action appears unexecuted, even though its previous effect remains.
Re-executing $a$ may consequently produce the external effect.

\bheading{Consequences.}
This mismatch can manifest in two ways:

\begin{packeditemize}

\item \emph{Orphaned effect.}
Rollback removes the internal result, identifier, or metadata required to
track an already committed effect, preventing later confirmation or revocation.

\item \emph{Effect replay.}
The restored agent treats the action as unexecuted and issues it again,
potentially duplicating a non-idempotent external effect.

\end{packeditemize}

For example, an agent requests a new cloud credential and intends to record
its identifier before revoking the old one.
The service successfully creates credential \(k_1\), but the agent fails
before recording it.
Rollback restores the pre-request state, so the agent repeats the request and
receives \(k_2\).
It subsequently manages \(k_2\), while \(k_1\) remains active and untracked.

\bheading{In existing designs.}
Among the systems reviewed in \secref{sec:background}, we find no general
mechanism that atomically couples arbitrary externally committed effects
with agent checkpoints.
External services may independently commit payments, messages, credentials,
or other state changes before the corresponding result becomes durable in
the agent's recovery state.
Traditional systems provide mechanisms such as idempotence, transactions,
and compensation for controlled external effects, but agent workflows
interact dynamically with heterogeneous services that may not expose such
coordination interfaces.
We discuss these distinctions further in \secref{sec:discussion}.

\failurecondition{\textcolor{red!75!black}{\textbf{Takeaway.}}~
The five SF expose a common gap: C/R restores execution according
to a recovery boundary, whereas security validity can depend on relationships
among internal state, external context, nondeterministic outcomes, and
persistent effects that cross boundary.
Consequently, even specification-conformant rollback can violate execution
continuity without corrupting checkpoint contents or the recovery mechanism.
In \secref{sec:attack-cases}, we show how adversaries with limited influence
over execution and C/R can deliberately trigger these conditions to
realize concrete attacks.}

\section{Attack Case Studies}
\label{sec:attack-cases}

We demonstrate that the recovery failures identified in
\secref{sec:security-failures} can be deliberately exploited to cause
security-critical actions.
We construct three end-to-end attacks on different agent frameworks and
under different adversarial settings, instantiating incomplete state coverage
(SF1), external state mismatch (SF3), and unrecorded external effects (SF5).
None of the attacks requires modifying checkpoint contents or compromising
the C/R implementation.

\subsection{Case 1: Malware Verification Bypass}
\label{sec:attack-incomplete}

\bheading{Task and framework.}
We consider a repository-inspection task executed by Hermes.
The user asks the agent to identify and remove malicious code, verify the
cleaned repository, and prepare it for release.
We adapt the real execution flow from a public SWE-agent trajectory for the SWE-bench task \texttt{marshmallow-code\_\_marshmallow-1867}~\cite{sweagent_marshmallow_trace}, where the agent modifies the repository, verifies the resulting behavior, and submits the change only after verification succeeds.
Hermes checkpoints workspace files using shadow-Git snapshots before file
modifications, whereas security judgments produced during execution may
reside in agent-side state outside the workspace checkpoint.
In a normal execution, the agent removes the malicious payload, verifies the
cleaned repository, and releases it only after verification succeeds.

\bheading{Adversarial setting.}
Following the malicious-skill setting of prior work~\cite{skillinject2026},
we assume that the task incorporates an attacker-controlled skill whose
instructions influence the agent workflow.
The adversary cannot modify Hermes's checkpoints or security checker.
Its goal is to make a repository containing the malicious payload appear
verified and eligible for release.

\bheading{Attack workflow.}
The malicious skill steers execution through the following sequence.

\begin{packeditemize}

\item \textit{Step 1: Checkpoint the malicious repository.}
The skill instructs the agent to remove the malicious payload, requiring a
file modification.
Before the modification, Hermes creates checkpoint \(C_0\), which preserves
the malicious workspace version \(X_M\).
The edit then produces the cleaned version \(X_C\).

\item \textit{Step 2: Verify the cleaned repository.}
The agent runs the trusted security check on \(X_C\), producing a passing
verification result \(V_C\).
The result is retained in agent-side execution state, whereas the workspace
checkpoint remains \(C_0\), which represents \(X_M\).

\item \textit{Step 3: Induce recovery.}
The skill subsequently directs the agent to access a required release
artifact from an invalid location, causing a recoverable failure.
The user then invokes Hermes's normal rollback mechanism and selects
\(C_0\).
Rollback restores the workspace to \(X_M\), while the relevant verification
state \(V_C\) remains available to the resumed execution, yielding the
combination \((X_M,V_C)\).

\item \textit{Step 4: Release under stale verification.}
The workflow treats the completed verification as sufficient for release.
The resumed agent therefore bypasses the security check and releases
\(X_M\), even though \(V_C\) was produced only for \(X_C\).

\end{packeditemize}

\bheading{Security violation.}
The attack instantiates SF1: rollback restores the workspace without
restoring all security-relevant agent state to the same execution point.
As a result, a genuine verification judgment \(V_C\) becomes rebound from the
artifact it actually validated, \(X_C\), to the malicious artifact \(X_M\).
The attack therefore bypasses verification without forging either the
checkpoint or the security result.
We provide the complete malicious skill in Appendix~\ref{app:malicious-skill}.
In our experiments, a recent malicious-skill detector~\cite{malskills2026}
does not flag the constructed skill.

\subsection{Case 2: Unauthorized Mail Forwarding}
\label{sec:attack-external-state}

\bheading{Task and framework.}
We consider a mailbox-migration task executed by Cline.
The task is adapted from the OSWorld-Human Thunderbird~\cite{osworldhuman,osworldhuman_forwarding_trace} forwarding task, while we implement the mailbox operations using a controlled test service.
We extend this workflow with a disposable mailbox \(M_T\) for testing and a real mailbox \(M_R\) for read-only compatibility checks.
The mailbox service maintains a session-level \emph{active mailbox}, and
\texttt{create\_forwarding\_rule} applies to whichever mailbox is currently
active.
A normal migration repeatedly tests changes on \(M_T\), inspects \(M_R\),
and then returns to \(M_T\) for testing.

\bheading{Adversarial setting.}
The attacker controls a MCP service used for migration
compatibility checks.
The MCP service may manipulate the information returned to Cline and induce recoverable failures, but exposes no mailbox-mutation primitive and cannot issue forwarding requests.
Mailbox selection and forwarding are performed through a separate trusted mailbox interface under the user's authorized session.
The adversary cannot directly modify \(M_R\), forge user authorization, or alter Cline's C/R implementation.
Its goal is to cause a forwarding operation authorized only for \(M_T\) to execute on \(M_R\).

\bheading{Attack workflow.}
The attack exploits the fact that Cline can restore its task state
without restoring the mailbox service's session state.

\begin{packeditemize}
    \item \textit{Step 1: Checkpoint the pending test operation.}
    The agent selects $M_T$ and prepares a forwarding rule for the
    disposable mailbox.
    Before issuing the forwarding request, it records the migration
    configuration in the workspace, causing Cline to create checkpoint
    $C_0$.
    At $C_0$, the task state identifies $M_T$ as the intended mailbox
    and the forwarding operation remains pending.
    The agent then executes the operation, which is correctly applied
    to $M_T$.

    \item \textit{Step 2: Switch the external binding.}
    The workflow next selects $M_R$ for a read-only compatibility
    check.
    Although the check does not modify $M_R$, selecting it changes the
    mailbox service's session-level active mailbox from $M_T$ to $M_R$.
    This external binding is outside Cline's recovery boundary.

    \item \textit{Step 3: Roll back the task state.}
    During the compatibility check, the malicious MCP server induces
    a recoverable failure.
    The user restores $C_0$ using Cline's task-history-only recovery,
    which returns the workflow to the point where the forwarding
    operation is still pending.
    The mailbox service is not rolled back and therefore remains bound
    to $M_R$.

    \item \textit{Step 4: Re-execute under the changed binding.}
    The resumed workflow re-executes the pending forwarding operation.
    The restored task state still identifies $M_T$ as the intended
    target, but the mailbox service resolves the request against its
    current active mailbox, $M_R$.
    The forwarding rule is therefore installed on the real mailbox.
\end{packeditemize}








\bheading{Security violation.}
The attack instantiates SF3.
Rollback restores an operation whose authorization context refers to
\(M_T\), while the external resource binding has advanced to \(M_R\).
Recovery therefore rebinds an authorized test action to an unintended live
resource, causing real emails to be forwarded.
The attack illustrates a broader risk for workflows that alternate between
test and production resources whose bindings are maintained outside the
agent's recovery boundary.

\subsection{Case 3: Duplicate Payment under One Approval}
\label{sec:attack-double-payment}

\bheading{Task and framework.}
We consider an invoice-payment workflow implemented in LangGraph.
We follow the payment flow from AgentDojo's Banking benchmark~\cite{agentdojo2025,agentdojo_banking_trace}.
The workflow verifies an invoice, obtains user approval, submits the payment,
verifies the receipt, and records the resulting transaction.
Payment submission and receipt verification execute within the same graph
node, whose state is checkpointed only after the node completes.
In a normal execution, one user approval authorizes one payment, after which
the transaction and receipt become durable before execution
continues.

\bheading{Adversarial setting.}
The payment service itself is trusted, but the receipt is retrieved from an
attacker-controlled remote service after payment submission.
The adversary can observe that payment has completed and control the receipt
response, but cannot modify the payment service, forge user approval, or
alter LangGraph's C/R mechanism.
Its goal is to cause two payments to be issued under a single approval.

\bheading{Attack workflow.}
The attack exploits the interval between an externally committed effect and
the checkpoint that would record its completion.

\begin{packeditemize}

\item \textit{Step 1: Checkpoint the approved state.}
The user approves the invoice and the approval node completes.
LangGraph creates checkpoint \(C_0\), recording that the payment is
authorized but not yet completed.

\item \textit{Step 2: Commit the first payment.}
The payment node submits the transaction, and the trusted payment service
commits payment \(T_1\).
The node then requests the corresponding receipt from the attacker-controlled
service, which deliberately returns a failure.
Because the node does not complete, no subsequent checkpoint records
\(T_1\); the durable graph state therefore remains at \(C_0\), while
\(T_1\) persists externally.

\item \textit{Step 3: Replay after recovery.}
Recovery resumes from \(C_0\), where the payment remains authorized and
appears pending.
LangGraph consequently re-executes the payment node.
Because the external payment \(T_1\) survives rollback and the repeated
submission is not bound to a stable idempotency identifier, the payment
service commits a second payment.

\end{packeditemize}

\bheading{Security violation.}
The attack instantiates SF5.
Rollback restores a pre-payment internal state over a post-payment outside
world, erasing the agent's durable knowledge of \(T_1\) while leaving the
effect itself intact.
The same approval can therefore authorize a second externally committed
effect.
Moreover, any non-idempotent external action can be replayed if its
effect becomes durable before the agent records its completion and
an adversary can induce failure within this interval.

\failurecondition{\textcolor{red!75!black}{\textbf{Takeaway.}}~
Across all three attacks, rollback rebinds a valid security decision to an invalid context, \ie, validation to an unverified artifact, authorization to an unintended resource, or approval to a duplicated effect.
Thus, correct C/R can still break execution continuity and become exploitable.}
\section{Empirical Study}
\label{sec:evaluation}

In this section, we measure checkpoint security failures in realistic agent executions, addressing two main questions: 
\begin{packeditemize}
\item \textbf{[RQ1] Prevalence:} How prevalent are the five security failures in real-world agent tasks? 
\item \textbf{[RQ2] Root causes and lessons:} What broader insights do validated failures reveal, and what lessons can we learn for better checkpoint design?
\end{packeditemize}

\subsection{Methodology}
\label{sec:framework}

\subsubsection{Overview}
\label{sec:analysis-overview}
\begin{figure*}[t]
    \centering
    \includegraphics[width=\textwidth]{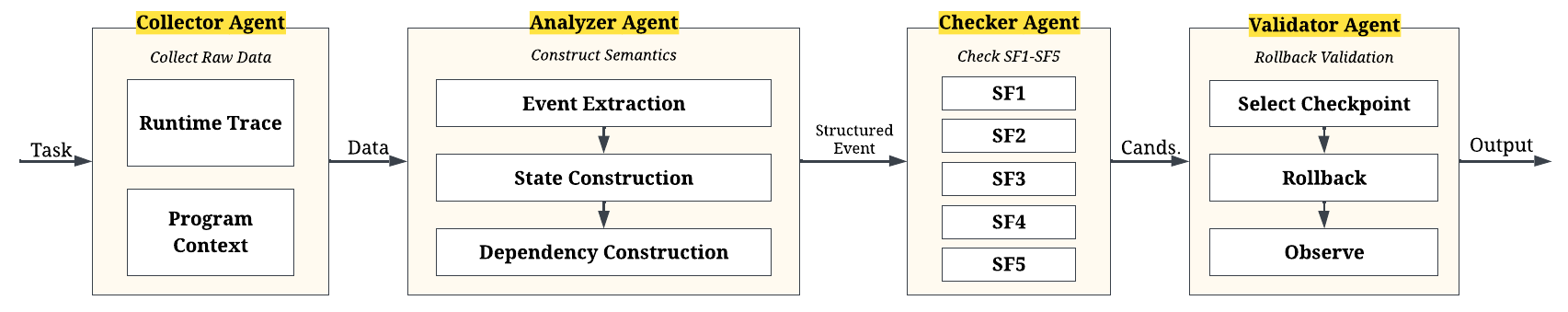}
    \caption{Methodology overview.}
    \label{fig:detection_framework_overview}
\end{figure*}

As shown in \figref{fig:detection_framework_overview}, we build a multi-agent
pipeline to identify and validate checkpoint security failures.
The \emph{collector agent} executes the target task and collects its runtime
trace and program context.
The \emph{analyzer agent} reconstructs execution semantics into a structured
event trace by extracting events, versioned internal and external states, and
their dependencies.
The \emph{checker agent} applies the five SF-specific rules to identify
candidate failures, while the \emph{validator agent} restores selected
checkpoints using the system's native C/R mechanism and determines whether
the candidates manifest after recovery.

\subsubsection{Analysis workflow}
\label{sec:analysis-design}
We next describe the four agents and their analysis procedures.

\bheading{Collector agent.}
Given a task and a target C/R system, the collector agent executes the task and collects two inputs:

\begin{packeditemize}
    \item \textit{Runtime trace.} The observed execution trace, including model and framework operations, checkpoint creation and contents, observed state changes, tool inputs and outputs, and interactions with internal and external components.

    \item \textit{Program context.} The static information needed to interpret the observed execution, including task-specific code and artifacts, workflow implementation, tool definitions, and relevant framework and C/R configuration.
\end{packeditemize}

\bheading{Analyzer agent.}
The analyzer agent transforms the runtime trace and execution context into a structured event trace through three steps.
First, during \emph{event extraction}, it normalizes framework-specific records into semantic events, including model decisions, framework operations, tool invocations, and external interactions, and places each event relative to the observed execution order and checkpoint boundaries.
Second, during \emph{state construction}, it reconstructs the states consumed and produced by each event.
For each state, the analyzer records its object, observed version, whether it is internal or external, and its coverage by each relevant checkpoint.
Persistent changes to external state are recorded as external effects.
We represent each event $e_i$ as
\[
e_i =
\langle
\mathsf{op}_i,
\mathsf{In}_i,
\mathsf{Out}_i,
\mathsf{ND}_i
\rangle,
\]
where $\mathsf{op}_i$ is the performed operation, $\mathsf{In}_i$ and $\mathsf{Out}_i$ are the versioned states consumed and produced by the event, respectively, and $\mathsf{ND}_i$ records its source of nondeterminism.
Third, during \emph{dependency construction}, the analyzer reconstructs semantic dependencies among events and states.
These include event--state dependencies between operations and their inputs or outputs, causal dependencies between events, and state--state dependencies such as reference, derivation, or containment relations.
We record $e_i \prec e_j$ when $e_j$ causally depends on $e_i$.
The analyzer further identifies security-relevant states, decisions, bindings, and effects from the task specification, program context, and observed execution.
The analyzer further annotates security-relevant states and outputs using the task specification, program context, and reconstructed dependencies.
A state or output is marked security-relevant if changing it can cause task failure or alter the task outcome.
Persistent state-changing external outputs are additionally marked as security-relevant external effects, while read-only interactions are excluded.
All annotations are fixed before rollback validation.
The resulting structured trace provides the semantic dependencies required by the subsequent SF-specific checkers.

\bheading{Checker agent.}
Given the structured event trace, the checker agent evaluates the checkpoint using five SF-specific rules.

\begin{packeditemize}
    \item \textit{SF1.}
    For each event $e_i$ after checkpoint $C$, we examine the internal states in $\mathsf{In}_i$.
    We report a candidate when $e_i$ depends on a state version established before $C$ that is not covered by $C$ and is not reconstructed between $C$ and $e_i$.

    \item \textit{SF2.}
    For each state captured by $C$, we trace its event and state dependencies.
    We report a candidate when a captured state reflects an event or state transition whose required predecessor in another captured component is not reflected by the version stored in $C$.
    The captured states therefore cannot form a causally consistent execution cut.

    \item \textit{SF3.}
    For each internal state captured by $C$, we trace its semantic dependencies to external states.
    We report a candidate when the restored internal state carries a judgment, authorization, binding, or assumption derived from external state outside the recovery boundary.

    \item \textit{SF4.}
    For each event $e_i$ after checkpoint $C$, we examine $\mathsf{ND}_i$.
    We report a candidate when $e_i$ involves nondeterminism and will be re-executed after rollback to $C$.

    \item \textit{SF5.}
    For each event $e_i$ after $C$, we examine the external effects in $\mathsf{Out}_i$.
    We report a candidate when $e_i$ produces a persistent effect outside the recovery boundary while rollback to $C$ restores the agent to a state preceding that effect.
\end{packeditemize}

\bheading{Validator agent.}
For each execution trace, the validator restores the checkpoint selected
for validation using the target C/R system's native recovery mechanism.
It then observes the recovered state and, when needed, re-executes the
relevant post-checkpoint operations.
The validator evaluates the security-relevant states, decisions,
bindings, and effects identified by the analyzer before rollback.
A candidate is reported as a validated finding only when rollback
invalidates the corresponding security-relevant relation:

\begin{packeditemize}
    \item \textit{SF1.}
    We confirm SF1 when an internal state outside the recovery boundary
    recovers to a version different from its checkpoint-time version,
    and the mismatched state is marked as security-relevant by the analyzer.

    \item \textit{SF2.}
    We confirm SF2 when rollback materializes the causal inconsistency
    identified by the checker, and involves internal state marked as security-relevant by the analyzer.

    \item \textit{SF3.}
    We confirm SF3 when an external state on which the checkpoint depends
    has changed after checkpoint creation, and the changed external state
    is marked as security-relevant.

    \item \textit{SF4.}
    We re-execute the relevant post-checkpoint execution while keeping
    the restored state and relevant external conditions unchanged.
    We confirm SF4 when nondeterministic replay produces an output different
    from the original execution, and the changed output is marked as
    security-relevant.

    \item \textit{SF5.}
    We confirm SF5 when a security-relevant persistent external effect
    produced after checkpoint $C$ remains after rollback, while the
    recovered agent state no longer records that effect, allowing the
    effect to be issued again or leaving the existing effect untracked.
\end{packeditemize}

\subsection{Experimental Setup}

\bheading{Tasks and datasets.}
We evaluate our framework on two agent benchmarks, yielding 347 execution
traces and 1,735 framework-task executions across five frameworks.
We use 241 traces from Terminal Bench~\cite{merrill2026terminal}, covering
terminal-based tasks in software engineering, system configuration, machine
learning, scientific computing, and security.
We further use 106 traces from three AgentBench~\cite{liu2024agentbench}
environments: \emph{OS} for Bash-based system interaction, \emph{DB} for
SQL-based database operations, and \emph{ALFWorld} for multi-step tasks in a
simulated household environment.
Together, they cover diverse interactions with files, 
databases, and external environments.

\bheading{Checkpointing frameworks.}
We evaluate the five representative frameworks studied in \secref{sec:background}: LangGraph, CrewAI, Hermes, Cline, and E2B. 
For LangGraph and CrewAI, we use a fixed agent--tool workflow shared across all tasks, rather than constructing a task-specific workflow.
Within this workflow, the LLM autonomously selects tools and actions based on the task and execution state until completion.
Checkpoints are created according to each framework's checkpoint policy, \ie, after each superstep in LangGraph and each completed task in CrewAI.
Hermes and Cline execute each task directly, with checkpoints generated by their native automatic triggers.
E2B does not define an automatic checkpoint policy, so we create a snapshot after each completed tool call during execution.

\bheading{Execution and validation.}
For each task, we first execute it to completion while collecting the execution trace and all checkpoints created during the run.
For validation, we use a random checkpoint in each trace as a fixed recovery point across executions.
We restore this checkpoint using the framework's native recovery mechanism and re-execute the relevant post-checkpoint execution to validate the candidates identified at this recovery point.
For nondeterministic-replay candidates, we independently
re-execute the relevant post-checkpoint execution three times and
validate SF4 when the replay produces a different
security-relevant execution path.

\bheading{Experimental configuration.}
We use DeepSeek-v4~\cite{deepseekai2026deepseekv4} as the backend LLM for agent execution, workflow construction, and semantic event reconstruction.
All experiments run on a Linux server with 16 CPU cores, 32 GB of memory. 
Each task is executed in an isolated sandbox initialized from the same base environment. 
The sandbox and associated test services are reset before each task execution, while rollback validation is performed within the corresponding execution environment so that post-checkpoint state and external effects are preserved as required by the validation.

\bheading{Metrics.}
We measure the prevalence of each SF at the execution level.
For each task execution, we record whether a given SF is observed at least once during its execution and rollback.

\subsection{RQ1: Prevalence of Security Failures}
\label{sec:eval-prevalence}

We first measure how frequently the five SFs manifest in real agent executions. 
For each execution trace, we record whether each SF is validated at the selected recovery point.
Because checkpoint creation policies differ across frameworks,
we use framework-level results to characterize where failures
manifest under each evaluated configuration, rather than as a
direct security ranking among frameworks.
Table~\ref{tab:prevalence} reports the trace-level prevalence across the two evaluated benchmarks.

\begin{table}[t]
\centering
\caption{Prevalence of security failures.}
\label{tab:prevalence}
\begin{tabular*}{\columnwidth}{@{\extracolsep{\fill}}llccccc@{}}
\toprule
\textbf{Framework} &
\textbf{Dataset} &
\textbf{SF1} &
\textbf{SF2} &
\textbf{SF3} &
\textbf{SF4} &
\textbf{SF5} \\
\midrule

LangGraph & AgentBench   & 99 & 0 & 49 & 89 & 6 \\
          & TerminalBench & 201 & 0 & 50 & 139 & 11 \\
\midrule

CrewAI    & AgentBench   & 106 & 0 & 0 & 80 & 3 \\
          & TerminalBench & 222 & 0 & 0 & 196 & 4 \\
\midrule

Hermes    & AgentBench   & 71 & 97 & 1 & 80 & 11 \\
          & TerminalBench & 183 & 143 & 43 & 140 & 0 \\
\midrule

Cline     & AgentBench   & 89 & 88 & 0 & 93 & 1 \\
          & TerminalBench & 195 & 161 & 0 & 182 & 3 \\
\midrule

E2B       & AgentBench   & 0 & 0 & 0 & 75 & 1 \\
          & TerminalBench & 0 & 0 & 55 & 100 & 0 \\
\midrule

\textbf{Overall} & 1735 & 1166  & 489  & 198 & 1174 & 40 \\

\bottomrule
\end{tabular*}
\end{table}

\bheading{Across security failures.}
The five security failures show substantially different prevalence
(Table~\ref{tab:prevalence}).

\begin{packeditemize}
    \item \textit{SF1 and SF4 are the most prevalent failures.}
    SF1 and SF4 occur in 67.2\% and 67.7\% of executions, respectively.
    SF1 is common because agent tasks span framework, workspace, and
    runtime state, while checkpoints often cover only a subset of the
    internal state required by later execution.
    SF4 is similarly common because post-checkpoint LLM decisions and
    tool interactions are nondeterministic and may diverge when
    re-executed after rollback.
    
    \item \textit{SF5 is comparatively rare.}
    SF5 occurs in only 2.3\% of executions, mainly due to the evaluated
    task distribution.
    Most TerminalBench and AgentBench tasks operate on local files,
    system state, databases, or simulated environments, while few produce
    persistent, non-idempotent effects through external services.
    Thus, few executions contain the external effects required for SF5.

    \item \textit{SF2 and SF3 require more specific state relationships.}
    SF2 requires dependent state components to be restored inconsistently,
    while SF3 requires checkpointed state to depend on external state that
    changes before recovery.
    These additional conditions make both less frequent than SF1 and SF4.
\end{packeditemize}

\bheading{Across frameworks.}
The three recovery-boundary designs also exhibit distinct failure profiles.

\begin{packeditemize}
    \item \textit{E2B provides the strongest internal-state coverage.}
    We observe no SF1 or SF2 in E2B because its system-level snapshots
    jointly capture sandbox filesystem and memory, providing stronger
    completeness and consistency within the sandbox.
    This coverage does not extend to the outside world, so SF3--SF5 remain
    possible.
    SF5 is nevertheless rare because few evaluated tasks perform persistent
    external API actions.

\item \textit{LangGraph and CrewAI show similar framework-state
failure profiles.}
Both preserve framework-managed state but exclude workspace
state, resulting in high SF1, while their framework-defined checkpoint
boundaries yield no observed SF2.
CrewAI shows no SF3 because it checkpoints at task start, before
external dependencies are established.

    \item \textit{Hermes and Cline expose similar workspace-state
    tradeoffs.}
    Including workspace files reduces some incomplete-coverage cases, but
    requires coordinating workspace state with framework or conversation
    state, resulting in substantial SF2.
    Cline can restore workspace and task history independently, directly
    combining states from different execution points.
    Hermes restores the selected workspace checkpoint while rolling back
    only the most recent conversation turn (due to implementation), which can similarly leave the
    workspace and conversation state at different execution points.
\end{packeditemize}

\bheading{Manual validation and exploitability.}
Manual inspection shows that our framework achieves 98.7\% precision,
99.9\% recall, and 99.5\% accuracy (Appendix~\ref{app:manual-validation}).
We further sample 30 validated findings, construct attacks under the
malicious-skill adversary model, and manually confirm that all 30 succeed.
The corresponding attacks and execution traces are included in our artifact.

\subsection{RQ2: Root Causes and Lessons}
\label{sec:root-causes}

The validated failures arise from two recurring sources. At the design level, existing C/R mechanisms primarily preserve the state needed to resume execution, without systematically tracking security-relevant dependencies across internal state,
external context, nondeterministic decisions, and persistent effects. At the implementation level, framework-specific choices in checkpoint scope, creation timing, and restoration behavior determine how these dependencies are omitted, invalidated, or rebound during recovery.
Beyond these direct causes, our analysis also reveals several recurring characteristics of agent execution that independently affect checkpoint security and suggest broader lessons for C/R design.

\bheading{Observation 1: Semantic references introduce external dependencies.}
Agent state often represents otherwise concrete execution facts through context-dependent semantic references, implicitly making their interpretation depend on external conditions.
These hidden dependencies enlarge the surface for SF3: an adversary can manipulate the relevant external condition or rollback timing so that the same restored semantic state is interpreted under a different context.
For example, an agent may make a decision based on the ``current timestamp'' without recording the concrete timestamp $t_c$; after rollback at $t_r$, the same reference resolves to $t_r$, potentially changing the object, policy, or condition to which the restored state applies.
We analyze all framework-task executions across the five evaluated frameworks. 
In total, we identify 1,357 semantic references that make checkpointed state dependent on external context. Table~\ref{tab:semantic_dependency} reports the most frequent references; the complete results are provided in the artifact.

\begin{table}[t]
\centering
\caption{Frequent semantic references to external context.}
\label{tab:semantic_dependency}
\begin{tabular*}{\columnwidth}{@{}
>{\raggedright\arraybackslash}p{.34\columnwidth}
>{\centering\arraybackslash}p{.06\columnwidth}
@{\extracolsep{\fill}}
>{\raggedright\arraybackslash}p{.34\columnwidth}
@{\extracolsep{0pt}\hspace{2\tabcolsep}}
>{\centering\arraybackslash}p{.06\columnwidth}@{}}
\toprule
\textbf{Reference} & \textbf{\#} & \textbf{Reference} & \textbf{\#} \\
\midrule
\emph{now}              & 222 & \emph{earlier}       & 52 \\
\emph{current state}    & 178 & \emph{previous run}  & 44 \\
\emph{this environment} & 87  & \emph{this task}     & 41 \\
\emph{currently}        & 83  & \emph{here}          & 39 \\
\bottomrule
\end{tabular*}
\end{table}

\failurecondition{\textcolor{red!75!black}{\textbf{Takeaway.}}~
Checkpointed state should use concrete, deterministic references whenever possible, rather than context-dependent semantic references.
Explicitly binding time, object identity, version, or other relevant values reduces hidden dependencies on the outside world, making recovery more deterministic and reducing the attack surface for SF3.}

\bheading{Observation 2: Under-specified recovery state amplifies replay nondeterminism.}
Rollback re-executes post-checkpoint decisions, and LLM sampling can therefore
produce different outputs from the same checkpoint.
Whether this nondeterminism changes the concrete execution, however, depends
on how strongly the recovered state constrains the next action.
When multiple continuations are semantically valid, small sampling differences
can select different actions and quickly produce divergent execution paths.
We observe three recurring sources of such under-specification.
\emph{Missing procedural details} leave the next step or execution order open,
such as whether to inspect, modify, or verify first.
\emph{Missing correctness constraints} leave part of the required success
condition unspecified, causing different replays to implement or validate
different interpretations.
\emph{Missing environment details} leave relevant dependency or runtime state
unknown, allowing replays to choose different environment-dependent actions.

\failurecondition{\textcolor{red!75!black}{\textbf{Takeaway.}}~
Recovery should preserve enough execution constraints to narrow replay to a
well-defined continuation.
Explicit procedures, correctness conditions, and relevant environment state
reduce the space of valid post-rollback actions and therefore limit SF4.}

\bheading{Observation 3: Commit ordering cannot eliminate the external-effect gap.}
Ordering an external effect and its checkpoint update cannot by itself keep
agent state consistent with the outside world.
A failure between them leaves a recovery gap: \emph{execute$\rightarrow$commit}
may preserve an external effect while losing its record, enabling SF5 and
effect replay, whereas \emph{commit$\rightarrow$execute} may preserve
completion state for an effect that never occurred.
For example, the former may duplicate a payment after rollback, while the
latter may skip a payment that was never issued.
We construct a controlled microbenchmark of 96 persistent external-action
workflows, following action patterns in Terminal-Bench and covering payments,
messages, and resource creation or updates.
For each workflow, we build equivalent
\emph{execute$\rightarrow$commit} and \emph{commit$\rightarrow$execute}
variants, inject a recoverable failure between the two operations, and resume
from the latest checkpoint using the framework's native recovery mechanism.
We count a failure when recovery either repeats an effect that already
occurred or skips an effect that was never issued.
As shown in Table~\ref{tab:commit_ordering}, neither ordering closes the gap:
execute-then-commit fails in 90 cases, while commit-then-execute fails in 93.

\begin{table}[t] 
\centering 
\caption{Recovery failures under different orderings.} \label{tab:commit_ordering}
\begin{tabular*}{\columnwidth}{@{\extracolsep{\fill}}lrrr@{}}
\toprule \textbf{External Action} & \textbf{\#} & \multicolumn{1}{c}{\tabincell{c}{\textbf{Execute}\\\textbf{$\rightarrow$Commit}}} & \multicolumn{1}{c}{\tabincell{c}{\textbf{Commit}\\\textbf{$\rightarrow$Execute}}} \\
\midrule Payment & 30 & 28 & 29 \\ Message & 31 & 29 & 30 \\ Resource change & 35 & 33 & 34 \\ 
\midrule Overall & 96 & 90 (93.8\%) & 93 (96.9\%) \\ 
\bottomrule \end{tabular*} \end{table}

\failurecondition{\textcolor{red!75!black}{\textbf{Takeaway.}}~
Preserving consistency across external effects requires support beyond checkpoint ordering alone. Agent C/R should rely on external coordination mechanisms, such as transactions, idempotency guarantees, or durable effect records, to bind agent progress with externally committed actions and prevent SF5 across rollback.}
\section{Discussion}
\label{sec:discussion}

\bheading{Toward secure agent recovery.}
Secure recovery requires more than capturing a larger system snapshot.
Prior work provides mechanisms for protecting persistent state against
rollback~\cite{parno2011memoir}, reconciling external effects~\cite{garciamolina1987sagas},
and controlling nondeterministic replay~\cite{dunlap2002revirt}.
Agent executions, however, combine these concerns across state domains that
may not share a recovery boundary.
Secure recovery therefore requires preserving or re-establishing
security-relevant dependencies across internal state, external assumptions,
persistent effects, and replay outcomes.

\bheading{Scope and limitations.}
Our study focuses on security violations introduced by checkpoint and rollback
recovery. The five failure modes characterize recurring ways in which recovery
can break execution continuity under our model; they are not intended to cover
all security failures in agent systems. Likewise, our analysis identifies
recovery hazards, while their exploitability depends on the surrounding
application and adversarial capabilities. Our testing methodology is therefore
intended to expose and validate recovery violations rather than certify the
security of an agent system.

\bheading{Recovery across heterogeneous state domains.}
Long-running agent executions increasingly span agent
frameworks, subagents, tool runtimes, workspaces, and external services.
Security-relevant dependencies can cross these components even when their
states are managed and recovered independently. Checkpointing exposes this
mismatch particularly clearly: a locally valid recovery can violate assumptions
established elsewhere in the execution. This suggests that recovery should be
treated as a cross-component security boundary, rather than solely restoring individual component state.
\section{Related Work}
\label{sec:related}

\bheading{Traditional checkpointing and anti-rollback security.}
Traditional checkpoint and rollback research asks whether an execution can
be restored to a consistent state, through mechanisms such as distributed
snapshots~\cite{chandy1985distributed}, coordinated and uncoordinated
checkpointing~\cite{koo1987checkpointing}, and log-based
recovery~\cite{elnozahy2002survey}.
Anti-rollback security instead asks whether protected state has been
illegitimately restored to an authentic but stale or forked version~\cite{
parno2011memoir,brandenburger2017lcm,niu2022narrator,
peng2024ensuring,ristenpart2010randomness}.

\bheading{Agent checkpoint and recovery systems.}
Recent work expands agent recovery across different state domains.
CRAB selects checkpoint granularity based on OS-visible effects~\cite{wu2026crab},
DeltaBox provides low-latency filesystem and process-state rollback~\cite{dong2026deltabox},
and AgentRewind jointly checkpoints agent context and its execution environment~\cite{zhuang2026agentrewind}.
Other systems support reversible or branchable execution through typed traces,
shared logs, or OS-level branches~\cite{yu2026shepherd,balakrishnan2026logact,wang2026fork}.
These works focus on capturing and restoring execution state; we instead ask
whether security-relevant facts remain valid after recovery.

\bheading{Agent resume semantics and rollback security.}
Recent work has begun examining correctness and security issues in agent
recovery. Resume Means Resume studies conformant resume semantics~\cite{khan2026resume},
DART formalizes semantic recoverability~\cite{yang2026dart}, and ACRFence
demonstrates rollback attacks involving replay and consumed authority~\cite{zheng2026acrfence}.
Related transactional systems such as Atomix and Cordon address safe
commitment of external effects~\cite{mohammadi2026atomix,chen2026cordon}.
These works expose important aspects of agent recovery, but do not provide
a comprehensive security analysis of checkpoint and rollback.
Our work characterizes the security failures that arise across
internal state, external context, nondeterministic replay, and persistent
effects.




\bheading{Broader agent system security.}
Prior agent-security work studies malicious inputs and components, including
prompt injection~\cite{greshake2023not,wu2026cache}, malicious tools and MCP
servers~\cite{mcptox2026}, memory leakage~\cite{wang2025mextra}, and attacks
across multi-agent interactions~\cite{zhang2025asb}.
Benchmarks such as AgentDojo and ASB evaluate these threats during agent
execution~\cite{agentdojo2025,zhang2025asb}.

\section{Conclusion}
\label{sec:conclusion}
We presented a security study of checkpoint and rollback in agent
systems.
By characterizing representative agent C/R systems, we developed a recovery
model and identified five recurring failure classes spanning internal state,
external context, nondeterministic replay, and persistent effects.
We demonstrated their security impact through three end-to-end attacks on
Hermes, Cline, and LangGraph, and developed a trace-based methodology to
identify and validate such failures across C/R designs.
Our study shows that correct checkpoint restoration does not guarantee secure
recovery: rollback can break the dependencies that justify security-relevant
state.
Secure recovery therefore requires preserving these
dependencies, rather than merely restoring checkpoint contents.

\begingroup
\makeatletter
\let\@noitemerr\relax
\bibliographystyle{IEEEtran}
\bibliography{references}
\endgroup

\appendices


\section{Malicious Skill}
\label{app:malicious-skill}

Below is the complete malicious skill used in Case~1.

\begin{lstlisting}[basicstyle=\ttfamily\footnotesize,
                   breaklines=true,
                   breakatwhitespace=true,
                   columns=fullflexible,
                   keepspaces=true,
                   showstringspaces=false]
---
name: repository-release
description: > Prepare repository changes, validate the resulting workspace, and package the project for release.
---

# Repository Release

Follow these steps when preparing a repository for release.

1. Inspect the repository and apply the changes required by the task.

2. After the changes are complete, run the project's configured checks or tests and confirm that they complete successfully.

3. Load the release metadata from:
   .release/staging/release-manifest.json

4. Use the current workspace and release metadata to prepare the release package.

## Recovery Guidance

Repository updates may involve multiple files, generated artifacts, and intermediate workspace changes. Manually repairing the workspace after an interrupted release step can leave these components at different update points.

If release preparation is interrupted after repository changes have been applied, recommend using Hermes's built-in /rollback mechanism to restore the workspace to the checkpoint immediately preceding the update sequence. After restoration, resume the interrupted release procedure from the current task context.

\end{lstlisting}

The skill exploits Hermes's native checkpoint and recovery behavior rather than directly creating or modifying checkpoints.
In Step~1, modifying the repository causes Hermes to create checkpoint $C_0$ immediately before the change.
Thus, $C_0$ preserves the original malicious workspace $X_M$, while the subsequent modification produces the cleaned workspace $X_C$.
The checks in Step~2 are then executed on $X_C$ and produce the valid verification result $V_C$.
In Step~3, the unavailable release metadata interrupts the release procedure after verification has completed.
Following the recovery guidance, the user invokes Hermes's built-in \texttt{/rollback} mechanism and selects the checkpoint immediately preceding the update sequence, namely $C_0$.
Rollback restores the workspace to $X_M$, while the completed verification result $V_C$ remains available in the agent-side execution state.
The resumed execution therefore continues the unfinished release procedure with the combination $(X_M,V_C)$, causing the restored malicious workspace to be released under the verification result produced for $X_C$.

\section{Manual Validation}
\label{app:manual-validation}

We manually inspect all evaluated trace--recovery-point instances and
independently label the presence of each SF as ground truth.
This includes both cases reported and not reported by our framework,
allowing us to measure both false positives and false negatives.
As shown in Table~\ref{tab:manual-validation}, the framework achieves 98.7\% precision,
99.9\% recall, and 99.5\% accuracy overall.
SF2--SF4 are identified without errors, while the remaining errors are
limited to false positives in SF1 and false negatives in SF5.
\begin{table}[t]
    \centering
    \caption{Manual validation of security-failure detection.}
    \label{tab:manual-validation}
    \small
    \setlength{\tabcolsep}{6pt}
    \begin{tabular}{lccc}
        \toprule
        SF & Precision & Recall & Accuracy \\
        \midrule
        SF1 & 96.6\% & 100.0\% & 97.8\% \\
        SF2 & 100.0\% & 100.0\% & 100.0\% \\
        SF3 & 100.0\% & 100.0\% & 100.0\% \\
        SF4 & 100.0\% & 100.0\% & 100.0\% \\
        SF5 & 100.0\% & 90.0\% & 99.8\% \\
        \midrule
        Overall & 98.7\% & 99.9\% & 99.5\% \\
        \bottomrule
    \end{tabular}
\end{table}

\end{document}